\documentstyle[12pt]{article}

\begin{document}

\title{Chiral Phase Transition at Strong Coupling in Lattice QCD}
\author{Dong Chen\\Department of Physics\\Columbia University}
\date{}
\maketitle

\def\thepage{CU--TP--700 \ \ \  hep-lat/9506004}
\thispagestyle{myheadings}

\begin{abstract}
  We study the chiral phase transition with staggered fermions on
the lattice at finite temperature in the strong coupling
limit.  The thermodynamic potential is derived in the large $d$
approximation where $d+1$ is the dimension of space time.  Our
calculation is simpler than the conventional method and leads to
a simple physical interpretation for the approximation scheme.
\end{abstract}

\section{Introduction}

  The chiral phase transition has been extensively studied in the
context of the large $d$ expansion at strong coupling.  The
mean field result at finite temperature shows a second order
phase transition \cite{Damgaard:84}--\cite{Petersson:86}.  The
flavor dependence and the critical exponents of the transition
have also been studied \cite{Petersson:92,Karsch:92}.  However,
in the conventional treatment of the large $d$ expansion, the
links in the temporal direction are treated quite differently
from those in the spatial directions, leading to a complicated
determinant calculation.  This more accurate treatment of the
temporal links raises a question about the overall consistency of
this application of strong coupling method to the QCD phase
transition.  In this paper, a better motivated and simpler
approach is presented.  The final result of this new approach
agrees with that of previous works to leading order in $d$.

  We start from the partition function of lattice QCD with
staggered fermions.  In section 2, we derive some basic
formulae relating the chiral condensate to the thermodynamic
potential.  In section 3, we take the strong coupling limit $g^2
\rightarrow \infty$.  After integrating out the links in the
spatial directions, we make a large $d$ expansion and derive the
thermodynamic potential in a saddle point approximation to the
integration over an intermediate variable.  In section 4, we
integrate over the time links and the fermion fields to
explicitly evaluate the thermodynamic potential.  The phase
transition is studied afterwards.  Our method is much simpler
than the conventional treatment, yet it gives the same result to
leading order in the $1/d$ expansion.
Finally, we give a physical interpretation of the calculation which is
quite obvious in this new approach.

\section{Basic Formulation}

  We consider lattice QCD with staggered fermions on a $d+1$
dimensional asymmetric lattice at finite temperature.  The
lattice size is $N_{s}\,^{d} \times N_{t}$ with $N_{t}$ kept
fixed and $N_{s} \rightarrow \infty$.  The lattice spacing is
$a_{s}$ in the spatial directions and $a_{t}$ in the temporal
direction.  The partition function with an external source
$\sigma(x)$ coupled to $\overline{\chi}\chi(x)$ is defined as
\begin{eqnarray}
	Z[\sigma] & = & \int [d U d \chi d \overline{\chi}]
\exp \{ - ( S_{G} + S_{F} ) + \sum_{x} \sigma (x)
	\overline{\chi}(x) \chi(x) \}\:,
\end{eqnarray}
where the gauge action $S_G$ and the fermion action $S_F$ are
\begin{eqnarray}
	S_{G} & = & \beta_{s} \sum_{\Box_{s}}\{1- \frac{1}{2N}
  Tr(U_{\Box_{s}} + U_{\Box_{s}}^{\dagger})\} \nonumber \\
    & & + \beta_{t} \sum_{\Box_{t}}\{1- \frac{1}{2N} Tr(U_{\Box_{t}}
  + U_{\Box_{t}}^{\dagger})\}\:, \\
	S_{F} & = & \frac{1}{2}\,\xi\sum_{x}
  \{ \overline{\chi} (x) U_{t}(x) \chi(x+\hat{t})
  - \overline{\chi} (x+\hat{t}) U_{t}(x)^{\dagger} \chi(x) \}
  \nonumber \\
	& & +\, \frac{1}{2} \sum_{x}
  \sum_{k=1}^{d}\eta_{k}(x)\{\overline{\chi} (x) U_{k}(x)
  \chi(x+\hat{k}) - \overline{\chi} (x+\hat{k})
  U_{k}(x)^{\dagger} \chi(x)\}\: .
\end{eqnarray}
In the above equations,
\begin{eqnarray}
	\beta_{s} = \frac{2N}{g_{s}^2}\, \frac{1}{\xi}\: ,
	\hspace{0.5in}
	\beta_{t} = \frac{2N}{g_{t}^2}\, \xi\: ,
	\hspace{0.5in}
	\xi = \frac{a_{s}}{a_{t}}\: ,
\end{eqnarray}
and $\eta_k(x) = (-1)^{x_1+\cdots+x_k}$ is the staggered phase
factor.  Define
\begin{eqnarray}
	Z[\sigma] = e^{W[\sigma]} \: .
\end{eqnarray}
The Legendre transform of the free energy $W[\sigma]$ is the
thermodynamic potential
\begin{eqnarray}
  \Gamma[\Phi] =  \sum_{x} \sigma(x) \Phi(x) - W[\sigma]\: ,
\end{eqnarray}
where
\begin{eqnarray}
  \Phi(x) = \frac{\delta\, W[\sigma]}{\delta\, \sigma(x)}\: ,
  \hspace{0.5in}
  \sigma(x) = \frac{\delta\, \Gamma[\Phi]}{\delta\, \Phi(x)}\: .
\end{eqnarray}
The normal chiral condensate is related to $\Phi(x)$ by
\begin{equation}
  	\langle \overline{\chi}(x)\chi(x) \rangle =
  \left. \frac{\delta\,W[\sigma]}{\delta\,\sigma(x)}
  \right|_{\sigma(x)=0} = \left. \Phi(x) \right|_{\sigma(x)=0} =
  \left. \Phi(x) \right|_{\frac{\delta\,\Gamma[\Phi]}
  {\delta\,\Phi(x)}=0}\: . \label{eq:chi}
\end{equation}
Thus the expectation value $\langle \overline{\chi}\chi \rangle$
equals the value of $\Phi$ at the stationary points of
the thermodynamic potential $\Gamma[\Phi]$.

\section{Large $d$ expansion at strong coupling}

  We work in the strong coupling limit, $g_{s} \rightarrow
\infty, g_{t} \rightarrow \infty $.  To leading order in
$\beta_s$ and $\beta_t$, we will neglect \(S_{G}\) in the
functional integral.  We will start by integrating out the gauge
links in the spatial directions and applying a self-consistent
large $d$ expansion to the fermion fields.  The leading quadratic
terms in the quark bilinear $\overline{\chi}\chi$ will be
linearized by introducing a Gaussian integral over an
intermediate scalar field $\lambda(x)$. We will then make a
saddle point approximation to the $\lambda(x)$ integral and
derive the thermodynamic potential in a simple form.  The
integrations over the time links and fermion fields are left to
the next section.

After integrating over all the \(U_k\)'s in the spatial
directions, we get
\begin{eqnarray}
	Z[\sigma] & = & \int[d U_t d \chi d \overline{\chi}]
    \exp\,\{
  \sum_{x}\sigma(x)\overline{\chi}(x)\chi(x)\nonumber\\
  & & +\frac{\xi}{2}\sum_{x}(\overline{\chi}(x)U_{t}(x)
  \chi(x+\hat{t})- \overline{\chi}(x+\hat{t})U_{t}^{\dagger}(x)
  \chi(x))\nonumber\\
  & & + \sum_{x} \sum_{k=1}^{d} F(x,x+\hat{k})\}\:,
  \label{eq:partition}
\end{eqnarray}
where
\begin{eqnarray}
  F(x,y) & = & \frac{1}{4 N}
    \overline{\chi}(x)\chi(x)\overline{\chi}(y)\chi(y)
   + \frac{1}{32 N^{2} (N-1)}
    (\overline{\chi}(x)\chi(x))^{2}
    (\overline{\chi}(y)\chi(y))^{2}\nonumber\\
  & &  +\cdots \; .
\end{eqnarray}
$F(x,y)$ is a polynomial of the composite fields
$\overline{\chi}(x) \chi(x) \overline{\chi}(y) \chi(y)$ and
\linebreak $\varepsilon^{i_1 i_2 \ldots i_N} \varepsilon^{j_1 j_2
\ldots j_N} \overline{\chi}^{i_1}(x) \overline{\chi}^{i_2}(x)
\ldots \overline{\chi}^{i_N}(x) \chi^{j_1}(y) \chi^{j_2}(y)
\ldots \chi^{j_N}(y)$ which is symmetrical between the two
spatially neighboring sites $x$ and $y$.  $N$ is the number of
colors.  This is a result from the $U_k$ integral and from the
fact that we are neglecting the gauge action at strong coupling.
The summation over $d$ neighbors for $F(x,y)$ in
equation~(\ref{eq:partition}) leads to a large $d$ expansion
\cite{Petersson:83}.  It turns out that
$\langle\overline{\chi}\chi\rangle$ is on the order of
$\sqrt{1/d}$ and the higher order terms in $F(x,y)$ are
suppressed by powers of $1/d$.  A naive justification for this
$1/d$ expansion would be the mean field approximation.  In that
approximation, there would be a common factor $d$ in front of all
terms in $F$ and the $1/d$ expansion would amount to a standard
loopwise expansion.  The actual derivation is similar but more
involved.  For our current calculation, it is sufficient to keep
only the leading term in $F(x,y)$.  Follow the standard treatment
in \cite{Petersson:86}, we define
\begin{eqnarray}
    V(x,x') =  \frac{1}{2d}\sum_{k}(\delta_{x+\hat{k},x'}
  +\delta_{x-\hat{k},x'}) \: ,
\end{eqnarray}
which satisfies the following equations
\begin{eqnarray}
    V(x,x') =  V(x',x)\:,\hspace{0.25in} \sum_{x} V(x,x')  =  1\:,
    \hspace{0.25in} \sum_{x} V(x,x')^{-1}  =  1\:.
\end{eqnarray}
We rewrite the leading term of $\sum F(x,y)$ in the exponent as
\begin{eqnarray}
    	& & \exp \{\frac{1}{4 N}\sum_{x}\sum_{k=1}^{d}
  \overline{\chi}(x)\chi(x) \overline{\chi}(x+\hat{k})
  \chi(x+\hat{k})\} \nonumber\\
	& = & \exp\{\frac{d}{4 N}\sum_{x,x'}
  \overline{\chi}(x)\chi(x) V(x,x')
  \overline{\chi}(x')\chi(x')\} \nonumber\\
	& = & \int[d \lambda] \exp\{- \frac{N}{d}\sum_{x,x'}
  \lambda(x) V(x,x')^{-1} \lambda(x')+\sum_{x} \lambda(x)
  \overline{\chi}(x)\chi(x)\}\:.
  \label{eq:v}
\end{eqnarray}
Substituting the above equation into eq~(\ref{eq:partition})
and making the change of variables
$\lambda(x) \rightarrow \lambda(x) - \sigma(x)$, we have
\begin{eqnarray}
    Z[\sigma]
    & = & \int[d \lambda d U_{t} d \chi d \overline{\chi}]\,
  \exp\,\{ \sum_{x} \lambda(x) \overline{\chi}(x)\chi(x)
  \nonumber \\
    & & +\, \frac{\xi}{2}\sum_{x}(\overline{\chi}(x)U_{t}(x)
  \chi(x+\hat{t})- \overline{\chi}(x+\hat{t})
  U_{t}^{\dagger}(x)\chi(x))\nonumber \\
    & & -\frac{N}{d}\sum_{x,x'}(\lambda(x)-\sigma(x))V(x,x')^{-1}
  (\lambda(x')-\sigma(x'))\, \} \\
    & = & \int[d \lambda]
  \exp\{-\frac{N}{d}\sum_{x,x'}(\lambda(x)-\sigma(x))V(x,x')^{-1}
  (\lambda(x')-\sigma(x')) \nonumber \\
    & & + \sum_{x} A(\lambda(x))\,\}\:,
\end{eqnarray}
where
\begin{eqnarray}
    & & \exp\{\sum_{x}A(\lambda(x))\}\nonumber \\
    & = & \int[d U_{t} d \chi d \overline{\chi}]
  \exp\,\{ \sum_{x} \lambda(x) \overline{\chi}(x) \chi(x)
  \nonumber \\
    & & +\, \frac{\xi}{2} \sum_{x}(\overline{\chi}(x)
  U_{t}(x)\chi(x+\hat{t})- \overline{\chi}(x+\hat{t})
  U_{t}^{\dagger}(x)\chi(x))\, \} \: .
\end{eqnarray}
We now derive the thermodynamic potential.  Make saddle point
approximation to the $\lambda(x)$ integral,
\begin{eqnarray}
    Z[\sigma] & = & e^{W[\sigma]} \nonumber\\
    & = & \exp\, \{-\frac{N}{d}\sum_{x,x'}(\lambda(x)-\sigma(x))
  V(x,x')^{-1} (\lambda(x')-\sigma(x'))\nonumber\\
    & & + \sum_{x} A(\lambda(x)) \, \} \: . \label{eq:saddle}
\end{eqnarray}
The saddle point condition is
\begin{equation}
    - \frac{2N}{d}\sum_{x'}V(x,x')^{-1}(\lambda(x')-\sigma(x'))
  +A'(\lambda(x))= 0\:.
\end{equation}
We find
\begin{eqnarray}
    \Phi(x) & = & \frac{\delta W[\sigma] }{\delta \sigma}
  = A'(\lambda(x))\:,\\
    \Gamma[\Phi] & = & \sum_{x}\sigma(x) \Phi(x) - W[\sigma]
  \nonumber\\
    & = & - \frac{d}{4 N} \sum_{x,x'} \Phi(x) V(x,x') \Phi(y)+
  \sum_{x}B(\Phi(x))\:.
\end{eqnarray}
And $B(\Phi(x))$ is the Legendre transform of $A(\lambda(x))$
\begin{eqnarray}
    B(\Phi(x)) & = & \lambda(x) \Phi(x) - A(\lambda(x)) \:.
\end{eqnarray}

\section{Mean Field Results}
In this section, we integrate out the time links and the fermion
fields to obtain $A(\lambda)$ explicitly.  We then study the
thermodynamic function $\Gamma[\Phi]$.  We choose a different
integration order from the conventional treatment and give a
simpler derivation.

To calculate $A(\lambda)$, we will first integrate out the
$U_{t}$'s, keeping only the leading term in the
$\overline{\chi}\chi$.  Then we will introduce a scalar field
$\lambda_{t}(x)$ in order to rewrite the
$\overline{\chi}\chi\overline{\chi}\chi$ term into a bilinear
form.  Finally we will integrate over the Grassmann fields $\chi
\overline{\chi}$.
\begin{eqnarray}
    & & \exp \{\sum_{x} A(\lambda(x)) \} \nonumber \\
    & = & \int [d \chi d \overline{\chi}] \exp\,\{ \sum_{x}
    \lambda(x)
  \overline{\chi}(x) \chi(x)
  + \frac{\xi^{2}}{4N} \sum_{x} \overline{\chi}(x)
  \chi(x) \overline{\chi}(x+\hat{t}) \chi(x+\hat{t})\,\}
  \nonumber \\
    & = & \int [d \chi d \overline{\chi}] \exp\,\{ \sum_{x}
  \lambda(x)
  \overline{\chi}(x) \chi(x)
  + \frac{\xi^{2}}{4N} \sum_{x,x'} \overline{\chi}(x) \chi(x)
  V_{t}(x,x') \overline{\chi}(x') \chi(x') \,\} \nonumber \\
    & = & \int [d \lambda_{t} d \chi d \overline{\chi}] \exp\,\{
  - \frac{N}{\xi^{2}} \sum_{x,x'} \lambda_{t}(x) V_{t}^{-1}(x,x')
  \lambda_{t}(x') \nonumber \\
    & & + \sum_{x} (\lambda(x) + \lambda_{t}(x))\,
  \overline{\chi}(x) \chi(x) \,\} \nonumber \\
    & = & \int [d \lambda_{t}] \exp \{ - \frac{N}{\xi^{2}}
  \sum_{x,x'} \lambda_{t}(x) V_{t}^{-1}(x,x') \lambda_{t}(x') \}
  \times \prod_{x}\,(\lambda(x) + \lambda_{t}(x))^{N} \: ,
\end{eqnarray}
where
\begin{equation}
    V_{t}(x,x') = 1/2 ( \delta_{x,x'+\hat{t}}
  + \delta_{x,x'-\hat{t}} )\: .
\end{equation}
Making the assumption that both $\lambda(x)$ and $\lambda_{t}(x)$
are independent of $t$, we get
\begin{equation}
    A(\lambda(\vec{x})) = \frac{1}{N_{t}} \log\left(\int d
  \lambda_{t}(\vec{x}) \exp \{- \frac{N N_{t}}{\xi^{2}}
  \lambda_{t}(\vec{x})^{2}\} \times
  (\lambda_{t}(\vec{x})+\lambda(\vec{x}))^{N N_{t}}\right)\:.
\end{equation}

In order to study the thermodynamic function $\Gamma[\Phi]$, we
shall find the Legendre transform of $A(\lambda)$.  Let us first
discuss two regions of values for $\lambda$ where we can make
simple approximations:

    1) If $\lambda$ is large, on the order of $\sqrt{d}$,
the leading term in $A(\lambda)$ is $\lambda^{N N_{t}}$.  So we have
\begin{eqnarray}
    A(\lambda) & = & N \log \lambda + const\:, \\
    \Phi & = & A^{\prime}(\lambda) = \frac{N}{\lambda}\:.
\end{eqnarray}
It is natural to set $\Phi$ to a constant at this stage when
we evaluate $\Gamma[\Phi]$, so that
\begin{eqnarray}
    \Gamma[\Phi] & = & - \frac{d}{4 N} \sum_{x,x'} \Phi(x) V(x,x')
    \Phi(y)+ \sum_{x} N \log(\Phi(x)) \nonumber \\
    & = & (- \frac{d}{4 N} \Phi^2 + N \log \Phi) N_s^d N_t\:, \\
    \langle \overline{\chi} \chi \rangle & = &
  \left. \Phi(x) \right|_{\frac{\delta\,\Gamma[\Phi]}
  {\delta\,\Phi(x)}=0}
  = N \sqrt{\frac{2}{d}}\: .
\end{eqnarray}
This is the known result giving chiral symmetry breaking at zero
temperature.  We see that both $N_t$ and $\xi$ have disappeared
from the final answer for $\langle \overline{\chi}\chi \rangle$,
which means we are essentially working with a $d+1$ dimensional
system, identical to the original system at $T=0$.  The chiral
condensate is proportional to $\sqrt{1/d}$, which justifies the
large $d$ expansion.  Chiral symmetry is always broken.

    2) If $\lambda$ is small, we can expand $A(\lambda)$ in a
Taylor series.
\begin{eqnarray}
    A(\lambda(\vec{x})) & = & \frac{1}{N_{t}} \log \{C (1
  + a_{2} \lambda^{2} + a_{4} \lambda^{4} + \ldots)\}
  \nonumber \\
    & = & \frac{1}{N_{t}} \{\log C+a_{2} \lambda^{2}
  +(a_{4}-\frac{a_{2}^{2}}{2}) \lambda^{4} + \ldots \}\: ,
\end{eqnarray}
where
\begin{eqnarray}
    C & = & \int d \lambda_{t} \exp \{- \frac{N N_{t}}{\xi^{2}}
  \lambda_{t}^{2}\} \lambda_{t}^{N N_{t}}\:, \nonumber \\
    a_{2} & = & \frac{1}{C} \left( \begin{array}{c} N Nt \\ 2
  \end{array} \right) \int d \lambda_{t}
  \exp [- \frac{N N_{t}}{\xi^{2}} \lambda_{t}^{2}]
  \lambda_{t}^{N N_{t} - 2} = \frac{(N N_{t})^{2}}{\xi^{2}}\:,
  \nonumber \\
    a_{4} & = & \frac{1}{C} \left( \begin{array}{c} N Nt \\ 4
  \end{array} \right) \int d \lambda_{t}
  \exp [- \frac{N N_{t}}{\xi^{2}} \lambda_{t}^{2}]
  \lambda_{t}^{N N_{t} - 4}  \nonumber \\
    & = & \frac{(N N_{t})^{3}(N N_{t} - 2)}{6 \xi^{4}}\:.
\end{eqnarray}
Neglecting the constant term, and taking \(N N_t \gg 1\) just
for simplicity, we get
\begin{eqnarray}
    A(\lambda(\vec{x})) & = & \frac{N^2 N_t}{\xi^2} \lambda^2 -
  \frac{N^4 N_t^3}{3 \xi^4} \lambda^4\:, \\
    B(\Phi(\vec{x})) & = & \frac{\xi^2}{4 N^2 N_t} \Phi^2 +
  \frac{\xi^4}{48 N^4 N_t} \Phi^4 \:.
\end{eqnarray}
Again we take $\Phi(\vec{x})$ to be a constant field in space,
\begin{equation}
    \Gamma[\Phi] = [(- \frac{d}{4 N} + \frac{\xi^2}{4 N^2 N_t})
  \Phi^2 + \frac{\xi^4}{48 N^4 N_t} \Phi^4] N_s^d N_t\:.
  \label{eq:Gamma}
\end{equation}
Normally, when $d$ is large, the coefficient of the $\Phi^2$
term in the above equation has a negative sign and we have
broken chiral symmetry.  When the temperature is raised, $\xi$
becomes large.  If $\xi$ is on the order of $\sqrt{d}$, the
coefficient can change sign and a second order phase transition
occurs.  Above the transition, $\langle \overline{\chi} \chi
\rangle = 0$ and chiral symmetry is restored.  The critical $\xi$
for the chiral phase transition is, to leading order in $1/d$,
\begin{equation}
    \xi_c = \frac{a_s}{a_t} = \sqrt{N N_t d}\:.
\end{equation}
And the transition temperature $T_c$ is
\begin{equation}
  T_c = \frac{1}{N_t a_t} = \frac{\xi_c}{N_t a_s} = \frac{1}{a_s}
\sqrt{\frac{N d}{N_t}}\:.
\end{equation}

In general, large $d$ and high temperature are two competing
factors which determine the phase of the system.  This amounts to
finding the true minimum of the effective potential in
eq~(\ref{eq:saddle}),
\begin{eqnarray}
    V_{eff}(\lambda) & = & \frac{N}{d} (\lambda(x)-\sigma(x))^2
    - A(\lambda(x))
    \: ,
\end{eqnarray}
in the saddle point approximation.  For a fixed $d$ at low
temperature, the system is in the chiral symmetry broken phase.
A second order phase transition to the chirally symmetric phase
occurs when the temperature is raised to the order of
$\sqrt{d}$.

Our result above agrees with the conventional calculation
although our method is different.  In the conventional treatment,
$A(\lambda(x))$ is calculated in the following order:  One first
integrates over the Grassmann variables $\chi \overline{\chi}$,
resulting in a determinant involving the links in the temporal
direction.  One then performs the integral over these $U_t$'s.
Thus one performs an exact integration of the time links after
one make the large $d$ expansion.  However, this final integral
is quite elaborate and seems unnecessary.  In this paper, we
treat the temporal and spatial links similarly in that they are
all integrated out at the beginning.  We then do the $\chi
\overline{\chi}$ integral and finally integrate over the
intermediate field $\lambda_t$.  We arrive at an $A(\lambda)$
that has the same structure as that given by the conventional
treatment,
\begin{eqnarray}
    A(\lambda) & = & C \log(a_0 + a_2 \lambda^2 + \cdots
    + \lambda^{N_t N}) \:.
\end{eqnarray}
The difference lies only in the higher order $1/d$ corrections to
the coefficients in the logarithm of $A(\lambda)$.  Also, instead
of using an effective potential which depends on the intermediate
integration field, we derive the thermodynamic potential which is
directly related to $\langle \overline{\chi} \chi \rangle$ via
eq~(\ref{eq:chi}).

\section{Discussion}

When $d$ is large, one normally expects zero temperature physics
because the thermal fluctuations tend to average out in this
limit.  In terms of QCD, this leads to the chiral symmetry broken
phase.  However, even in the large $d$ limit, a chiral phase
transition can occur if the temperature is high enough.  From
eq~(\ref{eq:Gamma}), there are two terms contributing to the
coefficient of the $\Phi^2$ term in the thermodynamic potential
$\Gamma[\Phi]$.  The first term comes from the $d$ neighboring
sites which tends to drive the system into the chiral symmetry
broken phase, and the second term comes from the thermal
fluctuations in the Euclidean time direction.  When the
temperature is high enough, on the order of $\sqrt{d}$, the
thermal fluctuations from a single time direction will be
comparable to the total effect from all $2d$ spatially
neighboring sites and the chiral phase transition occurs.  In the
conventional strong coupling calculation,  the role of the
thermal fluctuation in the time direction is somewhat hidden
which leads to an unnecessarily complicated derivation.  In this
paper, a simpler and more unified approach is presented.  The
only difference in the treatment of the coupling between fermion
bilinears in spatial and temporal directions is that for the
spatial direction, the intermediate integration variables are
replaced by a constant determined from a saddle point condition;
while in the temporal direction, the corresponding quantity must
be explicitly integrated over to take the fluctuations into
account.  The chiral phase transition is second order in the
large $d$ limit at strong coupling.

\section{Acknowledgment}
  I wish to thank Professor Norman Christ for many helpful
discussions.

\end{document}